\documentclass[a4paper,12pt]{article}

\usepackage{amsmath}
\usepackage{amssymb}
\usepackage[dvips]{graphicx}
\usepackage[dvips]{psfrag}

\makeatletter
\@addtoreset{equation}{section}
\renewcommand{\theequation}{\thesection.\@arabic\c@equation}
\makeatother

\makeatletter
\renewcommand\appendix{\par
  \setcounter{section}{0}%
  \setcounter{subsection}{0}%
  \gdef\thesection{Appendix \@Alph\c@section }
  \renewcommand{\theequation}
  {\Alph{section}.\arabic{equation}}
}
\makeatother

\newcommand{\bracket}[1]{\left\langle #1\right\rangle}

\newcommand{\dd}{{\rm d}}
\newcommand{\ee}{{\rm e}}

\begin{document}

\titlepage

\vspace*{-15mm}   
\baselineskip 10pt   
\begin{flushright}   
\begin{tabular}{r}    
{\tt APCTP-Pre2009-008}\\
July 2009
\end{tabular}   
\end{flushright}   
\baselineskip 24pt   
\vglue 10mm   

\begin{center}
{\Large\bf
 Another Realization of Kerr/CFT Correspondence
}

\vspace{8mm}   

\baselineskip 18pt   

\renewcommand{\thefootnote}{\fnsymbol{footnote}}

Yoshinori~Matsuo\footnote[2]{ymatsuo@apctp.org}, 
Takuya~Tsukioka\footnote[3]{tsukioka@apctp.org} 
and 
Chul-Moon~Yoo\footnote[4]{c\_m\_yoo@apctp.org}

\renewcommand{\thefootnote}{\arabic{footnote}}
 
\vspace{5mm}   

{\it  
 Asia Pacific Center for Theoretical Physics, 
 Pohang, Gyeongbuk 790-784, Korea 
}
  
\vspace{10mm}   

\end{center}

\begin{abstract}
We study another realization of the Kerr/CFT correspondence. 
By imposing new asymptotic conditions for the near horizon geometry 
of Kerr black hole, 
an asymptotic symmetry which contains all of the exact isometries can 
be obtained. 
In particular, the Virasoro algebra can be realized as an enhancement of 
$SL(2, \mathbb R)$ symmetry of the AdS geometry. 
By using this asymptotic symmetry, 
we discuss finite temperature effects and show the correspondence 
concretely.  
\end{abstract}

\baselineskip 18pt   

\newpage

\section{Introduction}\label{sec:Introduction}

Recently, the correspondence between the Kerr black hole 
and conformal field theory was studied~\cite{ghss}.  
They investigated the near horizon geometry which 
had $SL(2, \mathbb R)\times U(1)$ isometries~\cite{baho},  
and considered the asymptotic symmetry 
following the work by Brown and Henneaux~\cite{bh}. 
The Virasoro algebra was realized from an enhancement of the 
rotational $U(1)$ isometry, not from the $SL(2, \mathbb R)$. 
It is natural to ask if there exists an another symmetry 
promoted from the $SL(2,\mathbb R)$ 
symmetry of the near horizon geometry. 
This symmetry is likely relevant to 
the deviation from the extremality. 

In this paper, we investigate another realization 
of an asymptotic symmetry. 
Imposing new asymptotic condition which is stronger than 
that in \cite{ghss}, we obtain an asymptotic symmetry  
which contains all of the exact isometries. 
The $SL(2,\mathbb R)$ isometry could be enhanced to the 
Virasoro algebra.  
We define the energy-momentum tensor 
following the studies on the quasi-local energy~\cite{by}. 
Then we calculate the finite temperature effect for 
the mass and the angular momentum by using this symmetry and 
show agreements with those 
of the near extremal Kerr black hole. 
The entropy can be also discussed by using the Cardy formula.

\section{Kerr Black Hole}\label{sec:Kerr}

We start by introducing the Kerr metric: 
\begin{align}
 \dd s^2 &= -\dd t^2 
  + \frac{2mr}{r^2 + a^2 \cos^2\theta} 
  \left(\dd t-a\sin^2\theta \dd\phi\right)^2 
  + \left(r^2 +a^2\right) \sin^2\theta \dd\phi^2 
 \notag\\&\quad
  + \frac{r^2 + a^2 \cos^2\theta}{r^2 -2mr +a^2} \dd r^2 
  + \left(r^2 + a^2\cos^2\theta\right)\dd\theta^2 , 
\end{align}
where we have used Boyer-Lindquist coordinate. 
The parameters $m$ and $a$ are related to the ADM mass 
and the angular momentum as 
\begin{align}
 M &= \frac{m}{G_N} , & 
 J &= \frac{am}{G_N} . 
\end{align}
The position of the horizon and the Hawking temperature are given by 
\begin{align}
 r_{\pm} &= m \pm \sqrt{m^2 -a^2} , &
 T_H &= \frac{r_+ - m}{4\pi m r_+} . 
\end{align}

We consider the near horizon geometry of the Kerr geometry. 
We define new coordinates 
\begin{align}
 t &= 2 \epsilon^{-1} a \hat t , &
 r &= a\left(1 + \epsilon \hat r \right) , &
 \phi &= \hat \phi + \frac{t}{2a},  \label{Rescale}
\end{align}
and take the limit of $\epsilon\to 0$ 
to obtain the near horizon geometry. 
For extremal case $a=m$, the near horizon geometry becomes  
\begin{equation}
 \dd s^2 = - f_0(\theta) \hat r^2 \dd\hat t^2 
  + f_0(\theta) \frac{\dd\hat r^2}{\hat r^2} 
  + f_\phi(\theta)\left(\dd\hat \phi + k \hat r \dd\hat t \right)^2 
  + f_\theta(\theta)\dd\theta^2 , 
\label{NearHorizon}
\end{equation}
with
\begin{align}
 f_0(\theta) &= f_\theta(\theta) = a^2 \left(1+\cos^2\theta\right) , &
 f_\phi(\theta) &= \frac{4a^2\sin^2\theta}{1+\cos^2\theta} , & 
 k &= 1 . 
\label{DetailOfGeometry}
\end{align}
Hereafter, we consider this near horizon geometry 
and omit `` $\hat\ $ '' of the coordinates. 
The near horizon geometry has $SL(2,\mathbb R)\times U(1)$ isometries  
generated by the following four Killing vectors:  
\begin{subequations}
\begin{align}
 \xi_{-1} &= \partial_t , & 
 \xi_0 &= t \partial_t - r \partial_r , & 
 \xi_{1} &= \left(t^2+\frac{1}{r^2}\right) \partial_t 
 - 2 t r \partial_r - \frac{2k}{r}\partial_\phi , \label{OriginalSL(2,R)}\\
 \xi_\phi &= \partial_\phi , 
\label{OriginalU(1)}
\end{align}
\end{subequations}
where $\xi_{-1}$, $\xi_0$ and $\xi_1$ form the $SL(2,\mathbb R)$, 
and $\xi_\phi$ is the $U(1)$ rotational symmetry.

\section{Asymptotic symmetry}\label{sec:symmetry}

Let us move on an asymptotic symmetry of 
the near horizon geometry of the Kerr metric. 
The asymptotic symmetry is defined by using 
the asymptotic boundary condition. 
The theory has an asymptotic symmetry 
if it has symmetry in the asymptotic region  
up to the small perturbations 
which satisfy the asymptotic boundary condition.
For geometries, asymptotic symmetries are  
specified by asymptotic Killing vectors 
which satisfy 
\begin{equation}
 \pounds_{\xi} g_{\mu\nu} = \mathcal O(\chi_{\mu\nu}),  
  \label{KillingEq}
\end{equation}
where $\pounds_\xi$ is the Lie derivative along $\xi$. 
Here, the asymptotic boundary condition is given by 
\begin{equation}
 h_{\mu\nu} = \mathcal O(\chi_{\mu\nu}) , 
\end{equation}
where $h_{\mu\nu}$ is a perturbation of the metric. 
We now impose the following boundary condition: 
\begin{equation}
 h_{\mu\nu} = 
  \bordermatrix{
  & t & r & \phi & \theta \cr
  t 
  & \mathcal O(r^{0}) 
  & \mathcal O(r^{-3})
  & \mathcal O(r^{-2}) 
  & \mathcal O(r^{-3}) 
  \cr 
  r 
  & 
  & \mathcal O(r^{-4})
  & \mathcal O(r^{-3})
  & \mathcal O(r^{-4})
  \cr 
  \phi 
  & 
  & 
  & \mathcal O(r^{-2})
  & \mathcal O(r^{-3})
  \cr 
  \theta
  & 
  & 
  & 
  & \mathcal O(r^{-3})
  }. 
\label{Constraints}
\end{equation}
The most general form of the asymptotic Killing vector 
which satisfies \eqref{KillingEq} is 
\begin{align}
 \xi &= 
  \left(
   \epsilon_\xi(t) + \frac{\epsilon_\xi''(t)}{2r^2} + \mathcal O(r^{-3})
  \right)
  \partial_t 
  + 
  \left(
   - r \epsilon_\xi'(t) + \frac{\epsilon_\xi'''(t)}{2r} + \mathcal O(r^{-2})
  \right)
  \partial_r 
 \notag\\&\quad
  + 
  \left(
   C - \frac{k \epsilon_\xi''(t)}{r} + \mathcal O(r^{-3})
  \right)
  \partial_\phi 
  + 
  \mathcal O(r^{-3})
  \partial_\theta , 
 \label{AsymptKilling}
\end{align}
where $\epsilon_\xi(t)$ is an arbitrary function of $t$, 
and $C$ is an arbitrary constant. 
We define $\xi_n$ as the asymptotic Killing 
with $\epsilon_\xi(t) = t^{1+n}$ and $C=0$ 
in (\ref{AsymptKilling}), 
since the Hawking temperature is zero for the extremal Kerr black hole 
(or equivalently time is non-compact).  
Then these vectors form the Virasoro algebra 
\begin{equation}
 [\xi_n, \xi_m]_{\text{LB}} = \pounds_{\xi_n}\xi_m = (m-n)\xi_{m+n} . 
\end{equation}

We should mention key differences from the original Kerr/CFT 
correspondence~\cite{ghss}. 
Our asymptotic condition \eqref{Constraints} 
is stronger than the original one. 
The constraint \eqref{KillingEq} can be divided 
into that from the background $\bar{g}_{\mu\nu}$ and the perturbation, 
$
 \pounds_\xi g_{\mu\nu} 
  = \pounds_\xi \bar g_{\mu\nu} + \pounds_\xi h_{\mu\nu} . 
$
The constraint from the perturbation is automatically 
satisfied when we impose the constraint from the background, 
differently from \cite{ghss} 
and similarly to \cite{bh}. 
Our asymptotic symmetry \eqref{AsymptKilling} contains 
all of the original symmetries 
\eqref{OriginalSL(2,R)} and \eqref{OriginalU(1)}. 
This is because our perturbation is small enough 
not to break the original symmetry asymptotically. 
Contrary, in~\cite{ghss}, 
the perturbation is so large such that 
it breaks original symmetry even asymptotically. 
The rotational Killing vector \eqref{OriginalU(1)} 
is just realized as $\xi$ with $\epsilon_\xi(t) = 0$ and $C=1$. 
It is crucial that 
the $SL(2,\mathbb R)$ Killing vectors \eqref{OriginalSL(2,R)} 
are identical to $\xi_n$ with $n=-1,\ 0,\ 1$. 
In our asymptotic symmetry, this $SL(2, \mathbb R)$ 
is enhanced to the Virasoro algebra.

\section{Asymptotic charge}\label{sec:charge}

We first consider a charge which is associated with 
the asymptotic symmetry~\cite{bb,bc}. 
The asymptotic charge is defined 
as the deviation of the charge from the background $\bar{g}_{\mu\nu}$. 
For infinitesimal perturbation $h_{\mu\nu}$, 
that is given by 
\begin{equation}
 Q^{\rm A}_\xi[h] 
  = \frac{1}{8\pi G_N}
  \!\int_{\partial\Sigma_\infty}\!\!\!\!\! 
  k_{\xi}[h,\bar g] , 
  \label{asymcharge}
\end{equation}
where the integration is taken over the boundary of a time slice.  
The two-form $k_\xi$ is defined by 
\begin{equation}
 k_{\xi}[h,\bar g] 
 = \frac{\sqrt{-\bar g}}{4}\epsilon_{\mu\nu\rho\sigma}
 \tilde k_{\xi}^{\mu\nu}[h,\bar g]\, \dd x^\rho \wedge \dd x^\sigma , 
\end{equation}
with
\begin{align}
 \tilde k_{\xi}^{\mu\nu}[h,\bar g]
 = \frac{1}{2}\Bigl[
& 
 \xi^\mu D^\nu h 
 - \xi^\mu D_\lambda h^{\lambda\nu} 
 + \left(D^\mu h^{\nu\lambda}\right)\xi_\lambda 
 + \frac{1}{2} h D^\mu \xi^\nu 
 \notag\\
&
 - h^{\mu\lambda}D_\lambda\xi^\nu 
 + \frac{1}{2}h^{\mu\lambda}
 \left(D^\nu\xi_\lambda+D_\lambda\xi^\nu\right) 
 - (\mu\leftrightarrow\nu) \Bigr] , 
\end{align}
where $D_\mu$ is a covariant derivative on the background geometry, 
and we denote $\bar g = \det \bar g_{\mu\nu}$ 
and $h = \bar g^{\mu\nu}h_{\mu\nu}$. 
The Dirac bracket of the asymptotic charge 
can be calculated by varying the charge as
\begin{align}
 \{Q^{\rm A}_{\zeta},Q^{\rm A}_{\xi}\} 
  &= \delta_{\zeta} Q^{\rm A}_{\xi} 
\nonumber 
 \\
 &= 
 \frac{1}{8\pi G_N}\!\int_{\partial\Sigma_\infty}\!\!\!\!\!
 k_{\xi}[\pounds_{\zeta}\bar g, \bar g]
 + \frac{1}{8\pi G_N}\!\int_{\partial\Sigma_\infty}\!\!\!\!\!
 k_{\xi}[\pounds_{\zeta} h , \bar g] . 
\end{align}
The second term is the charge $Q^{\rm A}_{[\zeta,\xi]}$, 
which is proportional to the perturbation $h$. 
The first term is an additional constant term, 
which provides the central charge of the algebra. 

For the near horizon geometry \eqref{NearHorizon}, 
we obtain 
\begin{subequations}\label{AsymptFlux}
 \begin{align}
 \tilde k_\zeta^{\mu r}[\pounds_\xi\bar g,\bar g] 
 &= 
 - \frac{k}{f_0(\theta)}\epsilon_\zeta(t)\epsilon_\xi'''(t) 
 + \mathcal O(r^{-2}) , & 
 &(\mu = \phi) 
 \\
 \tilde k_\zeta^{\mu r}[\pounds_\xi\bar g,\bar g] 
 &= \mathcal O(r^{-2}) . & 
 &\text{(other components)}
 \end{align}
\end{subequations}
Since 
we get the charge which depends on $t$, 
we consider the analytic continuation of $t$ and 
define the generators as\footnote{
This definition is an analogue of 
that for the Virasoro algebra in two-dimensional CFT i.e.\
$
 L_n = \frac{1}{2\pi i}\oint \dd z\ z^{1+n} T(z).  
$
We identify $Q_{\epsilon_n} \sim \epsilon_n(z)T(z)$ 
with $\epsilon_n(z) = z^{1+n}$. 
}
\begin{equation}
 L_n = \frac{1}{2\pi i}\oint \dd t\ Q^{\rm A}_{\xi_n} . 
\end{equation}
The central charge $c$ is then calculated as  
\begin{equation}
 \frac{c}{12}= 
  \frac{1}{2\pi i}\oint \dd t\ \frac{1}{8\pi G_N}
\!\int_{\partial\Sigma_\infty}\!\!\!\!\!
  k_{\xi_m}[\pounds_{\xi_n}\bar g, \bar g]=0,  
\end{equation}
namely,  
this algebra does not have the central extension within the framework 
to use the ``asymptotic charge'', and 
satisfies the commutation relation, 
\begin{equation}
 [L_n,L_m] = (n-m) L_{n+m} . 
\end{equation}

\section{Energy-momentum tensor in CFT}\label{sec:energy}

As shown in the previous section, 
the algebra of the asymptotic charge dose not have a central extension. 
This fact makes it difficult to discuss 
the correspondence of physical quantities between 
CFT and Kerr black holes. 
Hence, we use the ``quasi-local charge'' 
defined by using the surface energy 
momentum tensor~\cite{by}, and obtain anomalous transformations 
with a plausible cut-off. 

According to the GKPW relation~\cite{gkp,w} in 
the AdS/CFT correspondence~\cite{m}, 
an expectation value of energy-momentum tensor 
in the CFT is given by 
\begin{equation}
 \bracket{T^{\mu\nu}} 
  = \frac{2}{\sqrt{-\gamma}}
  \frac{\delta S_{\rm cl}}{\delta\gamma_{\mu\nu}} , 
\end{equation}
where $\gamma_{\mu\nu}$ is the metric on the boundary. 
This can be identified with 
the quasi-local energy-momentum tensor defined by Brown 
and York \cite{by} (see also \cite{bk}),  
\begin{equation}
 T^{\mu\nu}
  = \frac{2}{\sqrt{-\gamma}}
  \frac{\delta S_{\text{grav}}}{\delta\gamma_{\mu\nu}} . 
\end{equation}
The gravitational action $S_{\text{grav}}$ 
is the Einstein-Hilbert action 
with the Gibbons-Hawking term, 
\begin{equation}
 S_{\text{grav}} 
  = \frac{1}{16\pi G_N} \!\int_{\mathcal M}\!\! \dd^4 x\sqrt{-g} R 
  + \frac{1}{8\pi G_N}\!\int_{\partial\mathcal M}\!\!\!\! 
  \dd^3 x \sqrt{-\gamma} K, 
\end{equation}
where $K=K^\mu_\mu$ and $K_{\mu\nu}$ is the extrinsic curvature. 
Let us consider 
following decomposition of the metric: 
\begin{equation}
 \dd s^2 = N^2 \dd r^2 
+ \gamma_{\mu\nu}(\dd x^\mu + N^\mu \dd r)(\dd x^\nu + N^\nu \dd r). 
\end{equation}
We focus on the variation of the action with respect to 
the induced metric $\gamma_{\mu\nu}$. 
Since we only consider an on-shell action, 
the variation of the action is reduced to boundary terms, 
\begin{equation}
 \delta S = \!\int_{\partial\mathcal M}\!\!\!\! \dd^3 x\ 
  \pi^{\mu\nu} \delta\gamma_{\mu\nu} , 
\end{equation}
where $\pi^{\mu\nu}$ is the conjugate momentum of $\gamma_{\mu\nu}$. 
Then the energy-momentum tensor 
on the boundary is expressed as 
\begin{equation}
 T^{\mu\nu} 
= \frac{2}{\sqrt{-\gamma}}\,\pi^{\mu\nu} 
= \frac{1}{8\pi G_N} 
  \left(K^{\mu\nu}-\gamma^{\mu\nu}K\right) . 
\end{equation}
We next consider the ADM decomposition of the boundary 
metric $\gamma_{\mu\nu}$, 
\begin{equation}
 \dd s^2_{\partial\mathcal M} 
  = -N^2_{\partial\Sigma} \dd t^2 
  + \sigma_{ab} (\dd x^a + N^a_{\partial\Sigma} \dd t)
  (\dd x^b + N^b_{\partial\Sigma} \dd t) ,
\end{equation}
where $\Sigma$ is a time slice and 
$\partial\Sigma$ is that on the boundary. 
Then the quasi-local charge on the boundary can be defined by 
\begin{equation}
 Q^{\rm QL}_\xi = \!\int_{\partial\Sigma}\!\! 
\dd^2 x\sqrt{\sigma}\,
  u^\mu T_{\mu\nu} \xi^\nu , 
\label{QLCharge}
\end{equation}
where $u^\mu$ is a future pointing timelike 
unit normal to $\partial\Sigma$. 
This definition allows for the boundary $\partial\Sigma$ at 
a finite $r$. 
Here, we shall redefine 
the energy-momentum tensor on the boundary 
by its deviation from the background, 
\begin{equation}
 \tau^{\mu\nu}[h] = 
  T^{\mu\nu}\Bigr|_{g=\bar g + h} - T^{\mu\nu}\Bigr|_{g=\bar g} . 
\end{equation}
This redefinition is equivalent to introducing 
the boundary term to the action such that $\delta S = 0$ 
also at the boundary for $g_{\mu\nu}=\bar{g}_{\mu\nu}$. 
Then charges 
in the boundary theory 
become
\begin{equation}
 Q^{\rm QL}_\xi = \!\int_{\partial\Sigma}\!\! \dd^2 x\sqrt{\sigma}\,
  u^\mu \tau_{\mu\nu} \xi^\nu. \label{redefQLCharge}
\end{equation}
This definition of charges is slightly different from
that for the asymptotic charge (\ref{asymcharge})\footnote{
The relation between the quasi-local charge and the asymptotic charge 
is discussed in \ref{sec:chargerelation}. 
}.

Let us consider the anomalous transformation 
of the energy-momentum tensor which is given by 
\begin{equation}
 \delta_\xi T^{\mu\nu} = \tau^{\mu\nu}[\pounds_\xi \bar g]. 
\end{equation}
The mass and the angular momentum 
are defined as $M=Q_{\partial_t}^{\text{QL}}$ 
and $J=Q_{\partial_\phi}^{\text{QL}}$, respectively.%
\footnote{
In \cite{by}, the mass is defined by using 
$
 \xi^\mu = N_{\partial\Sigma} u^\mu , \label{BoundKilling}
$
instead of the Killing vector. 
However, this definition cannot give   
the ADM mass if the Killing vector is not 
orthogonal to the time slice. 
Hence we define the mass by using the Killing vector. 
}
We then obtain their anomalous transformations 
for the geometry \eqref{NearHorizon}, 
\begin{subequations}
\begin{align}
 \delta_\xi M &= - \frac{k^2}{8\pi G_N} \!\int \!\dd\phi\,\dd\theta\ 
  \frac{\left(f_\phi(\theta)\right)^{3/2}\sqrt{f_\theta(\theta)}}
  {2\Lambda f_0(\theta)} \epsilon_\xi'''(t) , \\ 
 \delta_\xi J &= - \frac{k}{8\pi G_N} \!\int\! \dd\phi\,\dd\theta\ 
  \frac{\left(f_\phi(\theta)\right)^{3/2}\sqrt{f_\theta(\theta)}}
  {2\Lambda^2 f_0(\theta)} \epsilon_\xi'''(t) , 
\end{align}
\end{subequations}
where we introduced a regularization by putting the boundary 
at $r = \Lambda$ with large but finite $\Lambda$. 
These terms vanish for $\Lambda\to\infty$. 
The regularization of $\Lambda$ can be 
interpreted as a cut-off in the field theory side. 
In the case of the Kerr black hole \eqref{DetailOfGeometry}, 
we obtain 
\begin{subequations}
\begin{align} 
 \delta_\xi M  
  &= - \frac{a^2}{G_N\Lambda} \epsilon_\xi'''(t) , 
  \label{AnomalyM} \\
 \delta_\xi J  
  &= - \frac{a^2}{G_N\Lambda^2} \epsilon_\xi'''(t) . 
  \label{AnomalyJ}
\end{align}
\end{subequations}

We assume this energy-momentum tensor 
corresponds to that of CFT, 
which obeys the Virasoro algebra coming  
from the transformation of $t$. 
We could define the generators as 
\begin{equation}
 L_n[h] = \frac{1}{2\pi i}\oint \dd t 
  \!\int_{\partial\Sigma}\!\!\dd^2x  
  \sqrt{\sigma}\, u^\mu \tau_{\mu\nu}[h]\,\xi_n^\nu(t) , 
  \label{BoundGenerator}
\end{equation}
where we have considered the analytic continuation of $t$. 
We have replaced the Killing vector by $\epsilon_\xi(t)\partial_t$, 
or equivalently picked up the leading term.       
The central charge can be read off from 
$L_n[\pounds_\xi\bar g]$ as 
\begin{equation}
 c = \frac{12 a^2}{G_N \Lambda} . \label{BoundCentral}
\end{equation}

\section{Correspondence}\label{sec:correspondence}

We now show the correspondence between 
the Kerr black hole and the boundary theory which is 
a (chiral) CFT. 
Charges in the boundary theory are identified 
with the asymptotic charges of the Kerr black hole. 
Since we take the Kerr geometry itself as the background, 
expectation values of charges should be zero at zero temperature. 
Hence, we consider finite temperature effects 
by using the conformal transformation  
from the zero temperature system. 

We first consider the analytic continuation of $t$ 
into a complex $z$ whose imaginary part is $t$. 
We map the complex plane of $z$ to cylinder 
by using the transformation, 
\begin{equation}
 z \to w = \frac{\beta}{2\pi}\log z , 
\end{equation}
where $\beta$ is the period of the imaginary part of $w$,  
so that we can obtain the finite temperature system with 
temperature $T=1/\beta$. 
The charges being zero at zero temperature, 
only anomalous effects have nontrivial contributions 
at finite temperature. 
Through the anomaly of mass \eqref{AnomalyM} 
and angular momentum \eqref{AnomalyJ}, 
we obtain the finite temperature effect, 
\begin{subequations}
\begin{align}
 M &= \frac{a^2}{2G_N\Lambda} (2\pi T)^2 , \label{BoundEnergy}\\
 J &= \frac{a^2}{2G_N\Lambda^2} (2\pi T)^2 . \label{BoundAngular}
\end{align}
\end{subequations}
An entropy in CFT can be calculated by using the Cardy formula~\cite{c}
which relates the central charge $c$ to the density of high-energy
states: 
\begin{equation}
 S = 2\pi \sqrt{\frac{c L_0}{6}} . 
\end{equation}
At finite temperature, $\epsilon_\xi$ should be expanded 
in Fourier modes $\ee^{-int/\beta}$, and 
$L_0$ corresponds to the energy. 
From \eqref{BoundCentral} and \eqref{BoundEnergy}, we obtain 
the entropy
\begin{equation}
 S = \frac{(2\pi)^2 a^2 T}{G_N \Lambda} . \label{BoundEntropy}
\end{equation}

In the gravity side, 
the finite temperature effects could be observed from the non-extremal case. 
In order to obtain the near horizon geometry with the AdS-like structure, 
we have to take the near extremal limit in which 
the difference between energy and angular momentum is 
infinitesimally small. 
We here consider the following relation: 
%
\begin{equation}
 m = a\left(1 + \epsilon^2\frac{\Delta}{2}\right), 
\end{equation}
where the parameter $\Delta$ implies the deviation from the 
extremality.   
Then the temperature is now given by 
\begin{equation}
 T_H = \epsilon\frac{\sqrt{\Delta}}{4\pi a} + \mathcal O(\epsilon^2).  
\end{equation}
The ADM mass and the angular momentum get the following corrections, 
respectively, 
\begin{subequations}
\begin{align}
 M &= \frac{a}{G_N}\left(1 + \epsilon^2\frac{\Delta}{2}\right), \\
 J &= \frac{a^2}{G_N}\left(1 + \epsilon^2\frac{\Delta}{2}\right) . 
\end{align}
\end{subequations}
The near horizon geometry becomes 
\begin{equation}
 \dd s^2 = - f_0(\theta) \left(r^2-\Delta\right) \dd t^2 
  + f_0(\theta) \frac{\dd r^2}{r^2-\Delta} 
  + f_\phi(\theta)\left(\dd\phi + k r \dd t \right)^2 
  + f_\theta(\theta)\dd\theta^2 , \label{FiniteTGeometry}
\end{equation}
where $f_0(\theta)$, $f_\phi(\theta)$, $f_\theta(\theta)$ and $k$ 
are same as in \eqref{DetailOfGeometry}. 
Due to the rescaling of time \eqref{Rescale}, 
the Hawking temperature of 
the near horizon geometry is 
\begin{equation}
 T_H = \frac{\sqrt{\Delta}}{2\pi} .  
\end{equation}
The mass of the near horizon geometry is defined by 
the charge associated to the timelike Killing $\partial_{\hat t}$ 
in the geometry, and is related to that of 
the original Kerr metric as 
\begin{equation}
 M_{(\text{near horizon})} 
  = 2a \epsilon^{-1} M_{(\text{original})} 
  - \epsilon^{-1} J_{(\text{original})} . 
\end{equation}
We consider the energy and the angular momentum 
given by their deviation from the extremality  
and obtain 
\begin{subequations}
\begin{align}
 M &= \epsilon \frac{a^2\Delta}{2G_N} 
 = \epsilon \frac{a^2}{2G_N}\left(2\pi T_H\right)^2 , \\
 J &= \epsilon^2\frac{a^2 \Delta}{2G_N} 
 = \epsilon^2 \frac{a^2}{2G_N}\left(2\pi T_H\right)^2 , 
\end{align}
\end{subequations}
where we wrote only the leading term in $\epsilon$. 
Therefore, we conclude that the mass and the angular momentum obtained 
in the CFT agree with 
those for the Kerr black hole, 
if we identify $\Lambda = \epsilon^{-1}$. 
The entropy of CFT also corresponds to its deviation from 
the extremality in the gravity side. 
The Bekenstein-Hawking entropy of the Kerr black hole is given by 
\begin{equation}
 S = \frac{2\pi m r_+}{G_N} 
  = \frac{2\pi a^2}{G_N}\left(1 + 2\pi \epsilon T_H 
			 + \mathcal O(\epsilon^2)\right) . 
\end{equation}
Its deviation from the extremality agrees with \eqref{BoundEntropy}. 

The relation between $\epsilon$ and $\Lambda$ can be 
understood as follows:  
The near horizon geometry comes from 
an infinitesimal region around the horizon. 
If the parameter $\epsilon$ in \eqref{Rescale} is kept finite, 
the geometry cannot be exactly equivalent to \eqref{NearHorizon} 
but can be approximated by \eqref{NearHorizon} 
in the near horizon region $r-r_+\ll \epsilon^{-1} a$. 
Since we cannot go out of this near horizon region 
as long as we use \eqref{NearHorizon}, 
the boundary of this geometry should be 
taken around $\hat r\lesssim\epsilon^{-1}$. 
Therefore we identify 
the position of boundary as $\Lambda = \epsilon^{-1}$.

\section{Conclusions and discussions}\label{conclusion}

We have considered another realization 
of the Kerr/CFT correspondence. 
We have constructed new asymptotic Killing vectors 
by imposing a stronger constraint in the asymptotic region 
of the near horizon geometry of the Kerr black hole. 
Perturbations which satisfy this constraint 
do not break the symmetry of the original geometry. 
Then our Killing vectors give an asymptotic symmetry 
which contains all of the exact isometries. 
These forms the Virasoro algebra which is 
an extension of the $SL(2,\mathbb R)$ symmetry of the geometry. 
We have calculated the central extension of this algebra 
by using the ``asymptotic charge''  
and it turns out to be zero. 
However this is not the end of the story. 
Indeed we found anomalies when we consider the ``quasi-local charge'',
instead. 

In order to see the anomalies in our case, 
we need some regularizations. 
Even if we apply the same regularization for the quasi-local charge 
into the asymptotic charge, we cannot obtain anomalous effects. 
Since the asymptotic charge is defined up to some asymptotic condition, 
it might not be appropriate to estimate
quantities which asymptotically vanish. 
In other words, the asymptotic charge should be 
defined at the boundary $r\to\infty$. 
On the other hand, 
the quasi-local charge can be defined at arbitrary $r$, 
and does not require any asymptotic condition. 
We could then introduce a regularization  
and obtain anomalies. 

Since the definition of the 
surface energy-momentum tensor 
has the same form to the GKPW relation, 
we have used it as the energy-momentum tensor in the CFT side. 
We have calculated finite temperature effects on 
the energy and angular momentum through their anomalies 
and shown the correspondence concretely.  
The entropy has been also derived by using the Cardy formula 
and has agreed with that from the extremal case. 
This fact is consistent in the following sense: 
Since charges are also defined as 
their deviation from the extremality, 
we can expect that the CFT which obeys our Virasoro algebra 
describes the deviation from the extremality. 

We have defined the energy-momentum tensor 
in the field theory side by its deviation from the background. 
This is related to the boundary condition, $\delta S$ on the boundary. 
Instead of this, one can introduce boundary counter terms 
related to the boundary condition of the background. 
If geometry approaches asymptotically to the background $\bar g$, 
the corresponding boundary counter terms are defined 
such that $\delta S = 0$ also on the boundary for $g = \bar g$. 
By using this counter term, we can obtain 
the same energy-momentum tensor 
without taking the deviation from the background. 
However, the explicit form of the counter terms are  
not derived in this paper. 
This is left for future studies. 

\vspace*{5mm}

\noindent
 {\large{\bf Acknowledgments}}

We would like to thank L.-M. Cao for useful discussion 
at the early stage of this work. 
This work is supported by YST program in APCTP.

\vspace*{2mm}

\appendix

\section{Asymptotic and quasi-local charges}
\label{sec:chargerelation}

We consider the relation between the asymptotic charge 
and the quasi-local energy. 
The asymptotic charge can be rewritten in the following form, 
\begin{equation}
 Q^{\rm A}_\xi = 
  \int_{\partial\Sigma_\infty}\!\!\!\!\!\!
 \dd^2 x \sqrt{-g}\, \tilde k^{tr}_{\xi} 
 = \lim_{r\rightarrow\infty}\int_{\partial\Sigma}\!\! 
 \dd^2 x \sqrt{\sigma}\, 
 u_\mu n_\nu \tilde k^{\mu\nu}_{\xi} , \label{AsymptCharge}
\end{equation}
where $n^\mu$ is an outward pointing unit normal 
to the boundary $\partial\Sigma$. 
Comparing \eqref{QLCharge} and \eqref{AsymptCharge}, 
we can read off the relation between 
the ``flux'' $\tilde k^{\mu\nu}_\xi$ and 
the surface energy-momentum tensor $T^{\mu\nu}$ as 
\begin{equation}
 \tilde k^{\mu\nu}_\xi n_\nu \Bigr|_{r\to\infty} 
  \sim T^{\mu\nu} \xi_\nu . 
  \label{ABRelation}
\end{equation}
This relation is analogous to the GKPW relation, 
in which the flux is related to the current in boundary theory. 
In the case of one-form gauge field, 
it has an expansion in the form of 
\begin{equation}
 A_\mu = A_\mu^{(0)} + r^{-a}\bracket{J_\mu} + \cdots , 
\end{equation}
where $A_\mu^{(0)}$ is related to the source in the boundary theory. 
Then the flux which goes though the boundary is related to 
the current in the boundary theory, 
\begin{equation}
 F_{\mu r} \sim  - \partial_r A_\mu 
  \sim a \bracket{J_\mu} r^{-a -1} . 
\end{equation}

Let us check this relation for the anomalous transformation 
of the ``flux'' $k_\xi$. 
We consider the variation of the metric under 
the asymptotic symmetry \eqref{AsymptKilling}, 
and take $\pounds_\xi \bar g$ as a perturbation. 
We define
\begin{equation}
 \tilde\kappa_\zeta^{\mu r}[\pounds_\xi\bar g] 
  = N^{-1} T^{\mu\nu}\zeta_\nu\Bigr|_{g=\bar g +\pounds_\xi\bar g} 
  - N^{-1} T^{\mu\nu}\zeta_\nu\Bigr|_{\bar g}. 
\end{equation}
Again, we consider its deviation from the background. 
Then, for the near horizon geometry \eqref{NearHorizon}, 
we obtain 
\begin{subequations}
 \begin{align}
  \tilde\kappa^{\mu r}_\zeta[\pounds_\xi \bar g] 
  &= 
  - \frac{k}{f_0(\theta)}\epsilon_\zeta(t)\epsilon_\xi'''(t) , &
  &(\mu = \phi) 
  \\
  \tilde\kappa^{\mu r}_\zeta[\pounds_\xi \bar g] 
  &= 0 . & 
  &(\text{other components})
 \end{align}
\end{subequations}
By using the identification of \eqref{ABRelation}, 
this expression agrees with \eqref{AsymptFlux}. 
Then, we can define a charge in analogy with 
the asymptotic charge as 
\begin{equation}
 \tilde Q^{\rm A}_\xi
  =\int_{\partial \Sigma}\!\!\dd^2x\sqrt{\sigma}u_\mu n_\nu 
  \tilde\kappa_\zeta^{\mu \nu}[h] . 
\end{equation}
However, this is not identical to $Q^{\rm QL}_\xi$. 
When we define the charge, 
we consider the deviation from the background. 
For the asymptotic charge we take the difference of 
the charge itself. 
On the other hand, we take the difference of 
the energy-momentum tensor for the quasi-local charge. 
Then, these two definitions give different charge, 
even though they are same 
if we do not take the background into account. 

\vspace*{2mm}


\end{document}